\documentclass{aa}
\usepackage{graphics}
\begin{document}
   \thesaurus{ 24    
              (02.03.4; 
		02.12.3; 
		03.13.6)} 
\title{Expansions for nearly Gaussian  distributions}

\author{S. Blinnikov \inst{1,2} \and R. Moessner \inst{2}}

   \offprints{S. Blinnikov}

\institute{
Institute for Theoretical and Experimental Physics,
   117259, Moscow, Russia; \\ 
Sternberg Astronomical Institute, 119899 Moscow, Russia
 \and Max-Planck-Institut f\"ur Astrophysik,
D-85740 Garching, Germany 
}

\date{Received 16 July 1997; accepted 10 November 1997}

\maketitle

\begin{abstract}

Various types of expansions in series of Cheby\-shev-Hermite
polynomials currently used in astrophysics for weakly non-normal
distributions are compared, namely the Gram-Charlier, Gauss-Hermite and
Edgeworth expansions.
It is shown that the Gram-Charlier series is most suspect because of
its poor convergence properties. The Gauss-Hermite expansion is
better but it has no intrinsic measure of accuracy. The best results
are achieved with the asymptotic Edgeworth expansion.
We draw attention to the form of this expansion found by Petrov
for arbitrary order of the asymptotic parameter and present a simple
algorithm realizing Petrov's prescription for the Edgeworth expansion.
The results are illustrated by examples similar to 
the problems arising when fitting spectral line
profiles of galaxies, supernovae, or other stars, and for the case of
approximating the probability distribution of  peculiar velocities
in the cosmic string model of structure formation.

\end{abstract}


\section{Introduction}
\label{intr}

The normal, or Gaussian, distribution plays a prominent role in
statistical problems in various fields
of astrophysics and general physics.
This is quite natural, since the sums of random variables tend
to a normal distribution when the quite general conditions of the
central limit theorem are satisfied. 
In many applications, to extract useful information on the underlying
physical processes, it is more interesting to measure the deviations
of a probability density function (hereafter PDF) from the normal
distribution than to prove that it is close to the Gaussian one.
This has been done for example  in the work on peculiar velocities 
and cosmic microwave background
anisotropies in various cosmological models
(Scherrer \& Bertschinger \cite{scherb}; Kofman et al. \cite{kofbgnd};
Moessner et al. \cite{moessnpb};
Juszkiewicz et al. \cite{romanetal}; Bernardeau \& Kofman \cite{bernkof};
Amendola \cite{amend}; Colombi \cite{colom}; Moessner \cite{moessn}; 
Ferreira et al. \cite{fermasilk}; Gaz\-ta\-\~naga et al. \cite{gazt}),
in analyzing the
velocity distributions and fine structure in elliptical galaxies
(Rix \& White \cite{rixwhi}; van der Marel \& Franx \cite{marfra};
Gerhard \cite{gerh}; Heyl et al. \cite{heylhs}), and  in studies of
large Reynolds
number turbulence (Tabeling et al.  \cite{tabzbmw}).
In this paper we present a unified approach to the formalism used
in those applications, illustrating it by examples of  similar problems
arising in cosmology and in the theory of supernova line spectra.

The first
investigation of slightly non-Gaussian distributions was
undertaken by Che\-byshev  in the middle of the 19th century,
who studied in detail a family of orthogonal polynomials
which form a natural basis for the expansions of these
distributions. A few years later the same polynomials were also investigated
 by Hermite and they are called Chebyshev-Hermite  or
simply Hermite polynomials now (their definition was first given
by Laplace, see  e.g. Encyclopaedia of Mathematics \cite{ME}).

There are several forms of expansions using  Hermite polynomials,
namely the Gram-Charlier, Gauss-Hermite and Edgeworth expansions.
We introduce the notation in Sect. \ref{nomnot} and use the
simple example of the  $\chi^2$ distribution for various 
degrees of freedom to illustrate the
properties  of  Gram-Charlier expansions in Sect. \ref{chisq}.
In subsequent sections the  $\chi^2$ distribution is used for
testing the other two expansions. We note  in Sect. \ref{fourier}
that the  Gram-Charlier series is just a Fourier expansion
which diverges in many situations of practical interest,
whereas the Gauss-Hermite series has much better convergence properties.
However, we point out that another
series, the so called Edgeworth expansion, is more useful in  many
applications even if it is divergent, since, first, it is directly
connected to the
moments and cumulants of a PDF (the property which is lost in the
Gauss-Hermite series) and, second, it is a true asymptotic
expansion, so that the error of the approximation is  controlled.

Some applications (e.g. Bernardeau \cite{bern92},\cite{bern94};
Moessner \cite{moessn}) involve 
cumulants of higher order. These require the use of
a correspondingly higher order Edgeworth series, but only the first few
terms of the series are given in standard references
(Cramer \cite{cramer}; Abramowitz \& Stegun \cite{abrsteg};
Juszkiewicz et al. \cite{romanetal}; Bernardeau \& Kofman \cite{bernkof}).
In Sect. \ref{petr} we popularize  a derivation of the Edgeworth
expansion due to Petrov (\cite{petrov}, \cite{petr1}, \cite{petr2}) for 
an arbitrary order of the asymptotic
parameter, and we present a slightly simpler and more straightforward
way of obtaining his result. The formula found by Petrov
(see Sect.~\ref{petr} below) requires a summation over indices with
non-trivial combinatorics which hindered its direct application.
We find a simple algorithm realizing Petrov's prescription for
any order of the  Edgeworth expansion and this algorithm is easily
coded, e.g. with standard Fortran, eliminating the need for symbolic
packages.

We find that the
Edgeworth-Petrov expansion is indeed very efficient and reliable.
In Sect. \ref{CS} we apply the formalism to the
problem of peculiar velocities resulting from  cosmic strings studied by
Moessner (\cite{moessn}) 
and we show how this technique allows one to reliably extract deviations
from Gaussianity, even when they are tiny.

\section{Background and notation}
\label{nomnot}

Let $F(x)$ be the cumulative probability distribution of a random 
variable $X$. Then the mean value for the random variable $g(X)$  
is the expectation
value
 \begin{equation}
      {\sf E}g(X)\equiv \langle g(X) \rangle\equiv
                     \int_{-\infty}^{\infty} g(x) dF(x) \; .
 \label{mean}
 \end{equation}
The PDF is $p(x)={dF(x)/dx}$. 
The distribution $F(x)$ is not necessarily smooth, so it can
happen that $p(x)$ is nonexistent
at certain points. Nevertheless, the mean ${\sf E}g(X)$
is defined as long as the integral in (\ref{mean})
exists. Following the definition (\ref{mean}), the $k$-th order moment
of $X$ is
 \begin{equation}
     \alpha_k \equiv {\sf E}X^k =\int_{-\infty}^{\infty} x^k dF(x) \; .
 \label{moment}
 \end{equation}
Thus, the mean of $X$ is $m \equiv \alpha_1 = {\sf E}X$ and its
dispersion is
 \begin{equation}
     \sigma^2 \equiv {\sf E}(X-{\sf E}X)^2=
                \int_{-\infty}^{\infty}(x-m)^2 dF(x) \; .
 \label{dispers}
 \end{equation}
We denote the cumulative normal distribution by 
 \begin{equation}
    P(x)\equiv\int_{-\infty}^x Z(t)dt
            ={1\over \sqrt{2\pi}}\int_{-\infty}^x\exp(-t^2/2)dt \; ,
 \label{normdist}
 \end{equation}
so its PDF is the Gaussian function
 \begin{equation}
    Z(x) = { \exp(-x^2/2) \over \sqrt{2\pi} } \; .
 \label{ggaus}
 \end{equation}

We also  need to recall  some definitions for sets of orthogonal
polynomials.
Any two polynomials $P_n(x)$ and $P_m(x)$ of degrees $n \neq m$
are orthogonal on the real axis with respect to the weight
function $w(x)$ if
 \begin{equation}
    \int_{-\infty}^\infty w(x)P_n(x)P_m(x) = 0 \; .
 \label{ortho}
 \end{equation}

We follow the notation of Abramowitz \& Stegun (\cite{abrsteg}) where
possible, so  $H\!e_n(x)$ denotes  the
polynomial with weight function $w(x)=\exp(-x^2/2) \propto Z(x)$.
According to  Rodrigues' formula,
 \begin{equation}
   H\!e_n(x) = (-1)^n e^{x^2/2} {d^n \over dx^n} e^{-x^2/2}  \; .
 \label{chherm}
 \end{equation}
We will refer to $H\!e_n(x)$
as  Chebyshev-Hermite polynomials following Kendall
(\cite{kend}).
The ones with the weight $w(x)=\exp(-x^2) \propto Z^2(x)$,
 \begin{equation}
   H_n(x) = (-1)^n e^{x^2} {d^n \over dx^n} e^{-x^2}  \; ,
 \label{herm}
 \end{equation}
will be called Hermite polynomials.

The expansions in Chebyshev-Hermite and Hermite polynomials
are used in many applications in astrophysics.
For a PDF $p(x)$ which is nearly Gaussian, it seems
natural to use the expansion
 \begin{equation}
   p(x) \sim \sum_{n=0}^{\infty} c_n {d^n Z(x)\over dx^n} \; ,
 \label{grtaylor}
 \end{equation}
where the coefficients $c_n$ measure the deviations of $p(x)$
from $Z(x)$. From the definitions (\ref{ggaus}) and (\ref{chherm}) it
 follows that
 \begin{equation}
    {d^n Z(x)\over dx^n} = (-1)^nH\!e_n(x)Z(x) \; ,
 \label{derg}
 \end{equation}
so that
 \begin{equation}
   p(x) \sim \sum_{n=0}^{\infty}(-1)^n c_nH\!e_n(x)Z(x) \; ,
 \label{grchp}
 \end{equation}
with
 \begin{equation}
   c_n = {(-1)^n \over n!}\int_{-\infty}^\infty p(t)H\!e_n(t) dt \; .
 \label{grcoeff}
 \end{equation}
This is the well-known Gram-Charlier series (of type A, see
e.g. Cramer \cite{cramer}; Kendall \cite{kend} and references therein 
to the original work).
Since $H\!e_n(x)$ is a polynomial, we see that the coefficient
$c_n$ in (\ref{grcoeff}) is a linear combination of the moments
$\alpha_k$ of the random variable $X$ with PDF $p(x)$. The combination
is easily
found by using the explicit expression for $H\!e_n(x)$,
which we derive in \ref{Lemmapl}, namely
 \begin{equation}
   H\!e_n(x) = n! \sum_{k=0}^{[n/2]}{(-1)^k x^{n-2k} \over
                        k! (n-2k)! \; 2^k }      \; .
 \label{chhermexpl}
 \end{equation}

The Gram-Charlier series was used in cosmological applications
in the paper by Scherrer \& Bertschinger (\cite{scherb}), but 
note that  it 
is incorrectly called   Edgeworth expansion there.
The Gram-Charlier series is merely a kind
of Fourier expansion in the set of polynomials $H\!e_n(x)$.
This expansion has poor convergence properties (Cram\'er 1957). 
Very often, for realistic cases, it diverges violently.
We consider such an example in the next section.
The Edgeworth series, discussed in detail in Sect. \ref{petr},
is a true asymptotic expansion of the PDF, which
allows one to control its accuracy.

\section{An example based on the $\chi^2$ distribution}
\label{chisq}

A good example for illustrating the fast divergence of the Gram-Charlier
series is given by its
application to the $\chi^2_\nu$ distribution with $\nu$ degrees  of freedom,
since the moments of this
distribution are known analytically and its PDF tends to the
Gaussian one for large $\nu$.
If $X_1$, $X_2$, ..., $X_\nu$, are independent,
normally distributed random variables with zero expectation and unit
dispersion, then the variable $\chi^2=\chi^2_\nu \equiv \sum_{i=1}^\nu X_i^2$
has the PDF
 \begin{equation}
   \rho(\chi^2)={(\chi^2)^{\nu/2-1}\exp(-\chi^2/2) \over
                   2^{\nu/2}\Gamma(\nu/2)},  \qquad \chi^2>0 \; .
 \label{chi2den}
 \end{equation}
The expectation value of $\chi^2$ is $\nu$ and its dispersion is $2\nu$.
If $x=(\chi^2-\nu)/\sqrt{2\nu}$, then $x$ has zero expectation and
unit dispersion and its PDF $p(x)$ asymptotically tends to
the Gaussian distribution $Z(x)$. Transforming $\rho(\chi^2)$ from
(\ref{chi2den}) to $p(x)$, we obtain  for $x>-\sqrt{\nu/2}$:
 \begin{equation}
   p(x)=\sqrt{2\nu}{(\sqrt{2\nu}x+\nu)^{\nu/2-1}\exp{-(\sqrt{2\nu}x+\nu)/2} \over
              2^{\nu/2}\Gamma(\nu/2)}  \; .
 \label{pxden}
 \end{equation}
The $\chi^2$  distribution was employed by 
Matsubara \& Yoko\-ya\-ma (\cite{matyok})
for a representation of the cosmological density field   (see also
Luo \cite{luo} for an application of the $\chi^2$  distribution in
the context of  cosmic microwave background temperature anisotropies).
It can also serve as a crude model for approximating the profiles of
spectral lines in moving media. If the lines are nearly Gaussian in the
matter at rest, then the motion with a high velocity gradient, like
that in supernova envelopes, leads to distortions which can
be approximated by the $\chi^2_\nu$ PDF (with $\nu=2$ for highly non-Gaussian
profiles, see e.g. Blinnikov \cite{blinnexp}, and references therein).

\begin{figure}
 \resizebox{\hsize}{\hsize}{\includegraphics{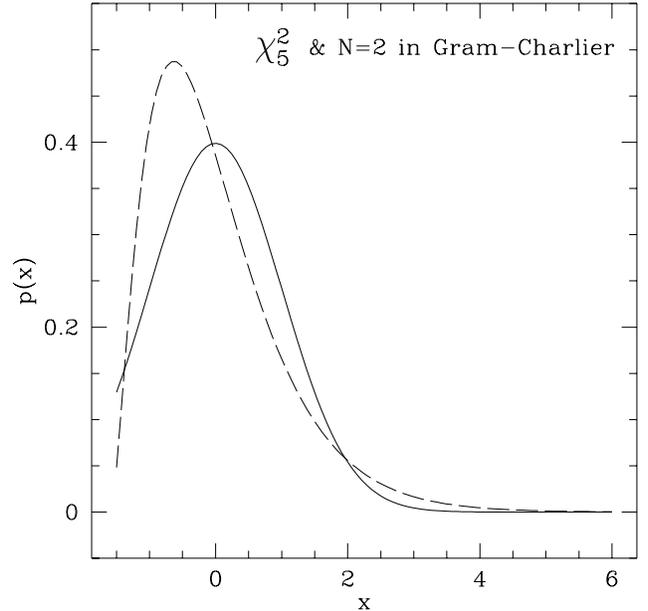}}
 \resizebox{\hsize}{\hsize}{\includegraphics{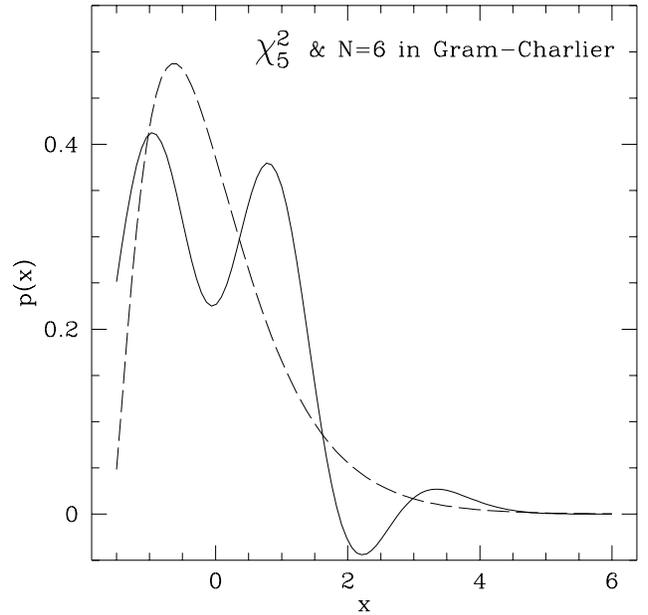}}
\caption{The normalized $\chi^2$ PDF
(\protect\ref{pxden}) for $\nu=5$ (dashed line), and
  its Gram-Charlier approximations
  with 2 and 6 terms in the expansion (solid line)}
\label{gr5-2-6}
\end{figure}

\begin{figure}
  \resizebox{\hsize}{\hsize}{\includegraphics{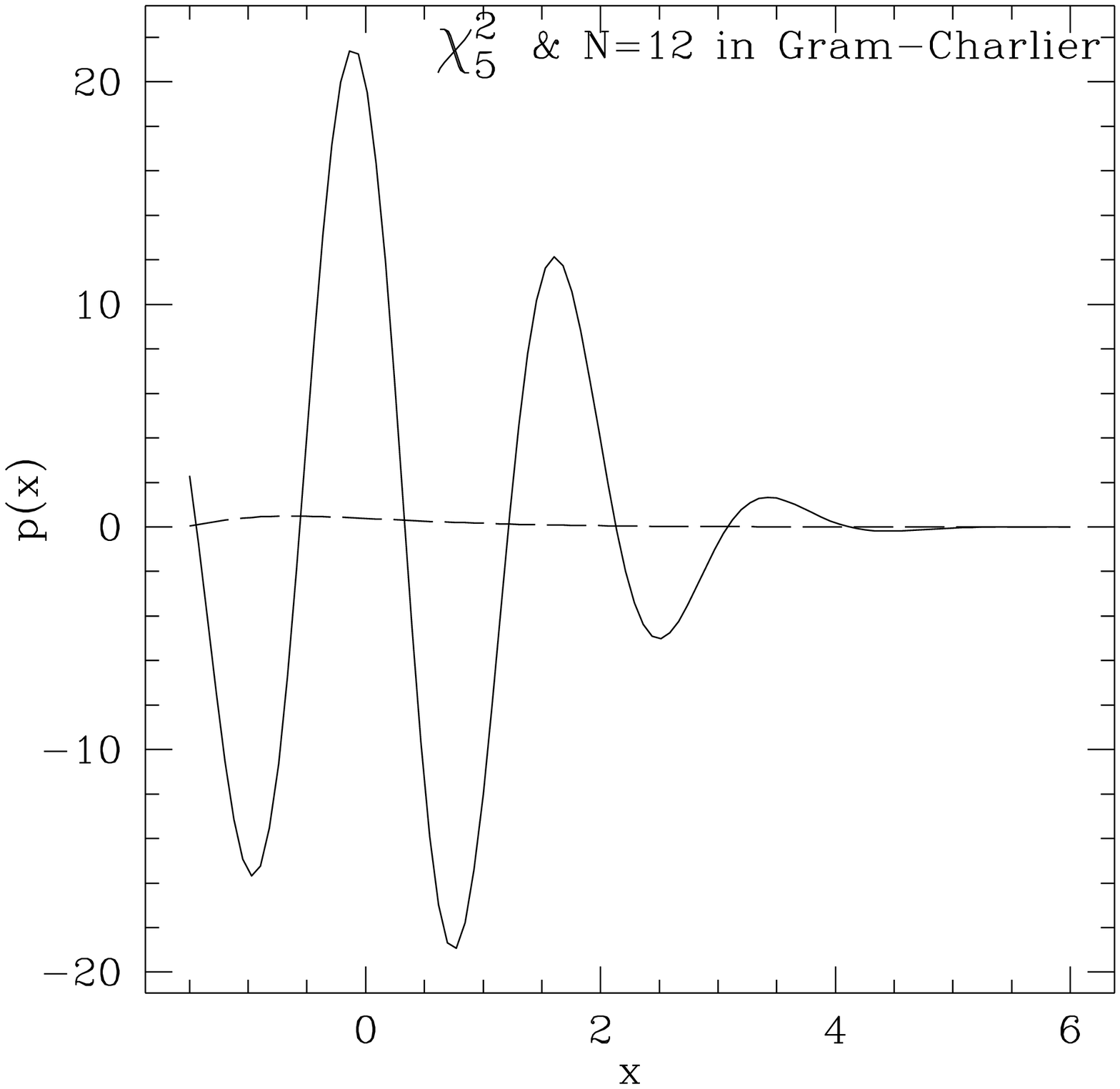}}
  \resizebox{\hsize}{\hsize}{\includegraphics{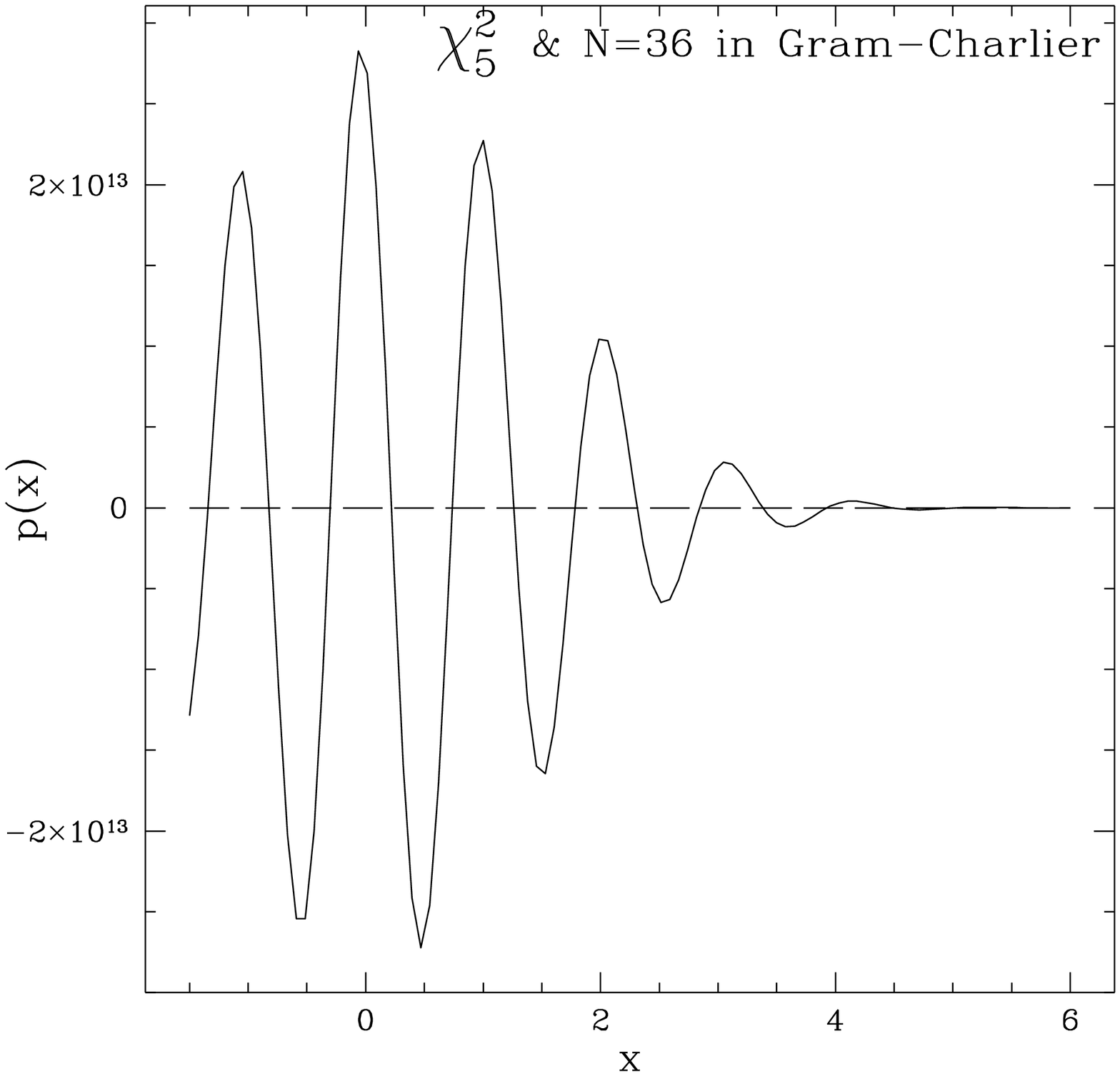}}
\caption{The same as in Fig. \protect\ref{gr5-2-6} but for
         12 and 36 terms in the Gram-Charlier expansion.}
\label{gr5-12-36}
\end{figure}

The comparison of the Gram-Charlier approximations of $p(x)$ with 
the exact results is presented
in Figs. \ref{gr5-2-6} and \ref{gr5-12-36} for an increasing number of terms
in the expansion. It is clear that the series quickly becomes
inaccurate with a larger number of terms included.


\section{Fourier expansions}
\label{fourier}


In order better to understand the poor convergence properties of
the Gram-Charlier series, let us first discuss how it is related
to  the Fourier expansion. A Fourier expansion (Szeg\"o
\cite{szegoe}; Suetin \cite{suetin}; Nikiforov \& Uvarov \cite{nikuvar})
for any function $f(x)$ in the set of orthogonal polynomials $P_n$ is
given by

 \begin{equation}
   f(x) \sim \sum_{n=0}^{\infty} a_nP_n(x) \; ,
 \label{fourp}
 \end{equation}
with
 \begin{equation}
   a_n = {1\over h_n}\int_{-\infty}^\infty w(t)f(t)P_n(t) dt \; .
 \label{coeff}
 \end{equation}
Here $h_n$ is the squared norm
 \begin{equation}
   h_n = \int_{-\infty}^\infty w(t)P_n^2(t) dt \; ,
 \label{normgen}
 \end{equation}
and
 \begin{equation}
   h_n=\sqrt{2\pi}n! \qquad {\rm for} \qquad H\!e_n(x) \; ,
 \label{normHC}
 \end{equation}
 \begin{equation}
   h_n=\sqrt{\pi}2^n n! \qquad {\rm for } \qquad H_n(x) \; .
 \label{normH}
 \end{equation}
Now we can see that the Gram-Charlier series (\ref{grchp}) is just
the Fourier expansion (\ref{fourp}) of $f(x)=p(x)/Z(x)$ in the set of
Chebyshev-Hermite polynomials with  $c_n=(-1)^n a_n$.

The properties of the Gram--Charlier approximations of $p(x)$
in Figs. \ref{gr5-2-6} and \ref{gr5-12-36} are to be considered
in the general context of the convergence of Fourier expansions.
The source of the divergence  lies in the sensitivity of
the Gram-Charlier series to the behavior of $p(x)$ at infinity --
the latter must fall to zero faster than $\exp(-x^2/4)$ for the
series to converge (Cramer \cite{cramer}; Kendall \cite{kend}). 
This is often too restrictive
for practical applications. Our example of the $\chi^2$ distribution in
(\ref{pxden}), with its exponential behavior at infinity, clearly
demonstrates this.

\begin{figure}
  \resizebox{\hsize}{\hsize}{\includegraphics{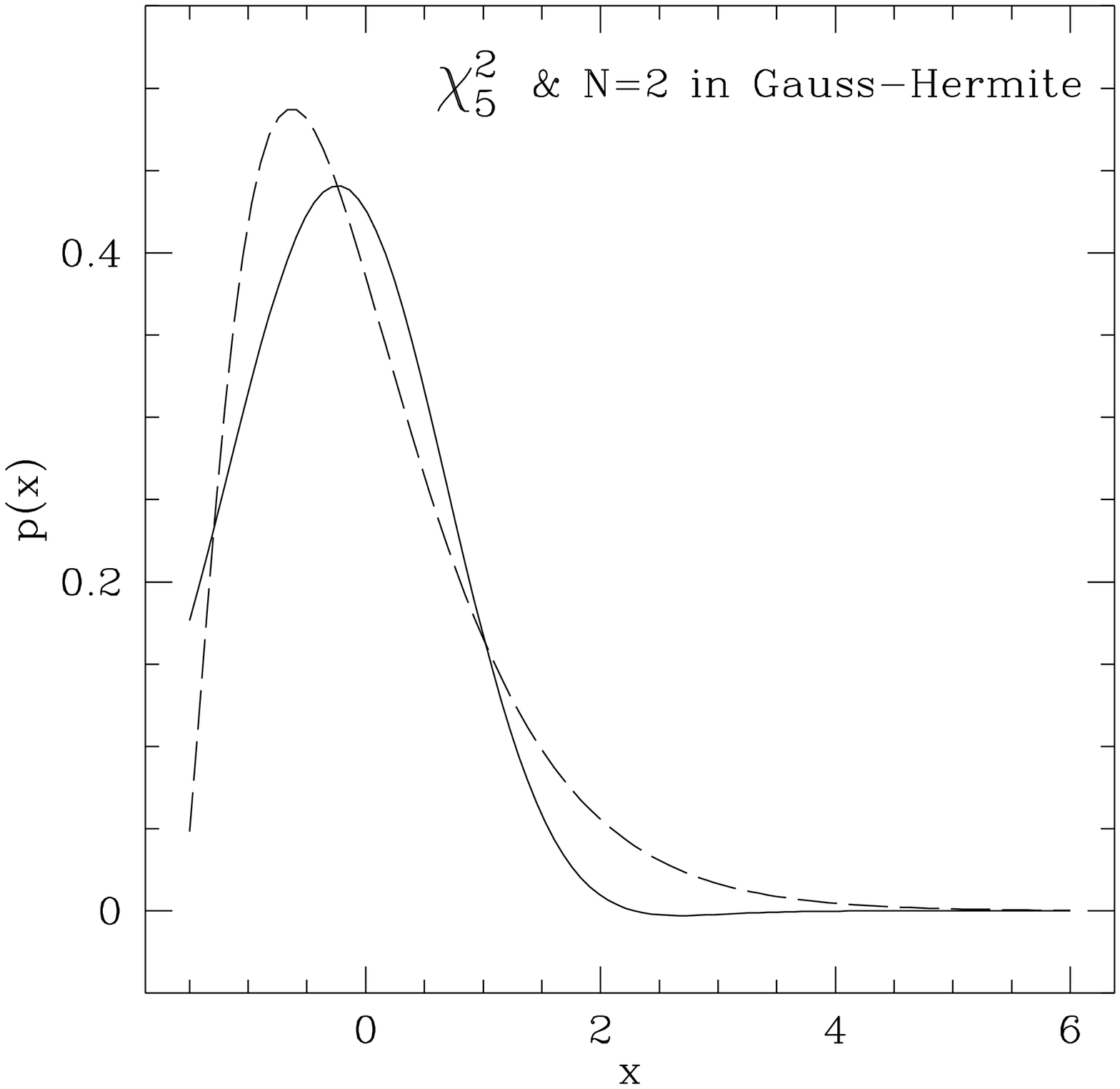}}
  \resizebox{\hsize}{\hsize}{\includegraphics{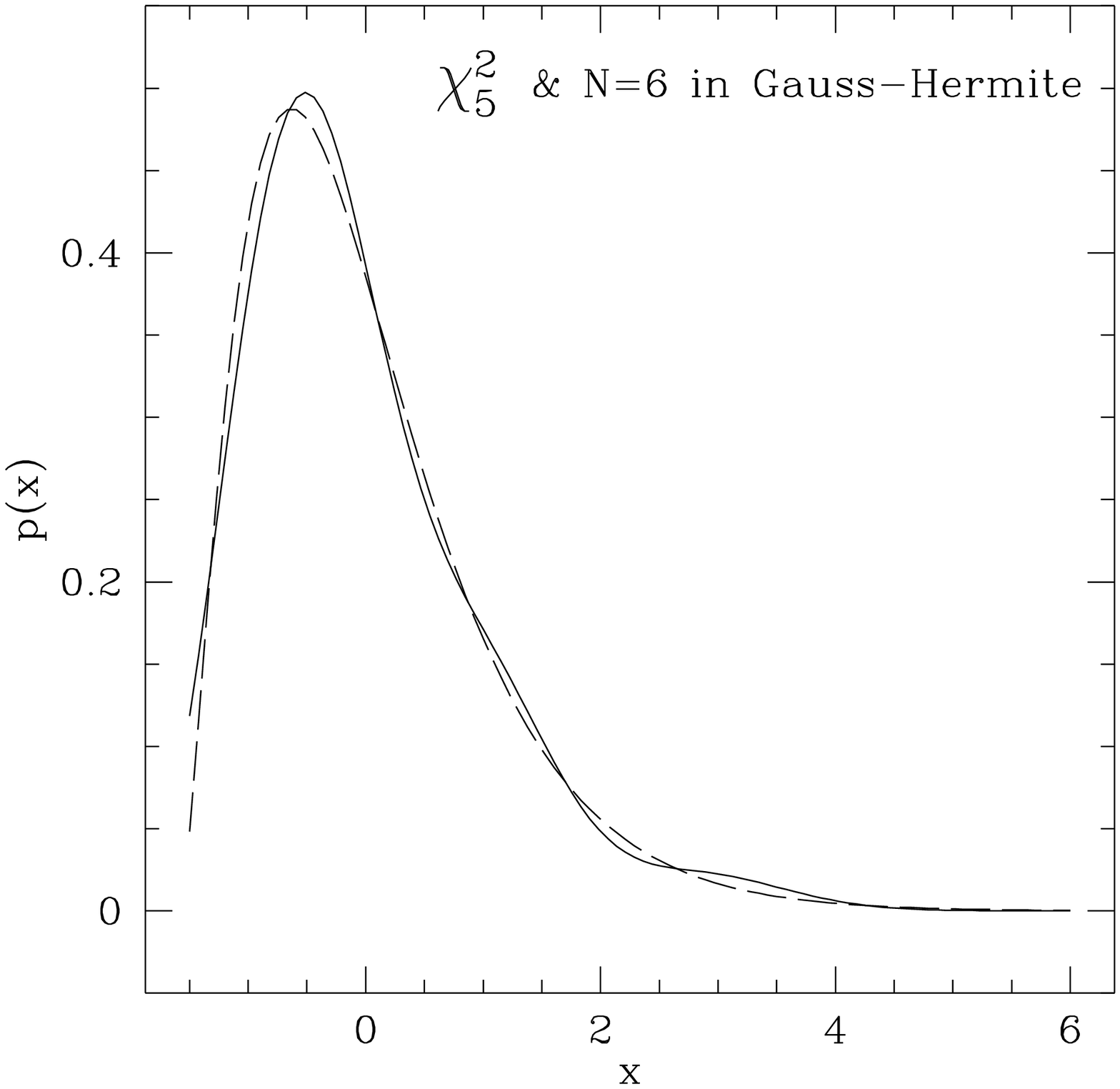}}
\caption{The normalized $\chi^2$ PDF
(\protect\ref{pxden}) for $\nu=5$ (dashed line), and
  its Gauss-Hermite approximations
  with 2 and 6 terms in the expansion (solid line).}
\label{gh5-2-6}
\end{figure}

\begin{figure}
  \resizebox{\hsize}{\hsize}{\includegraphics{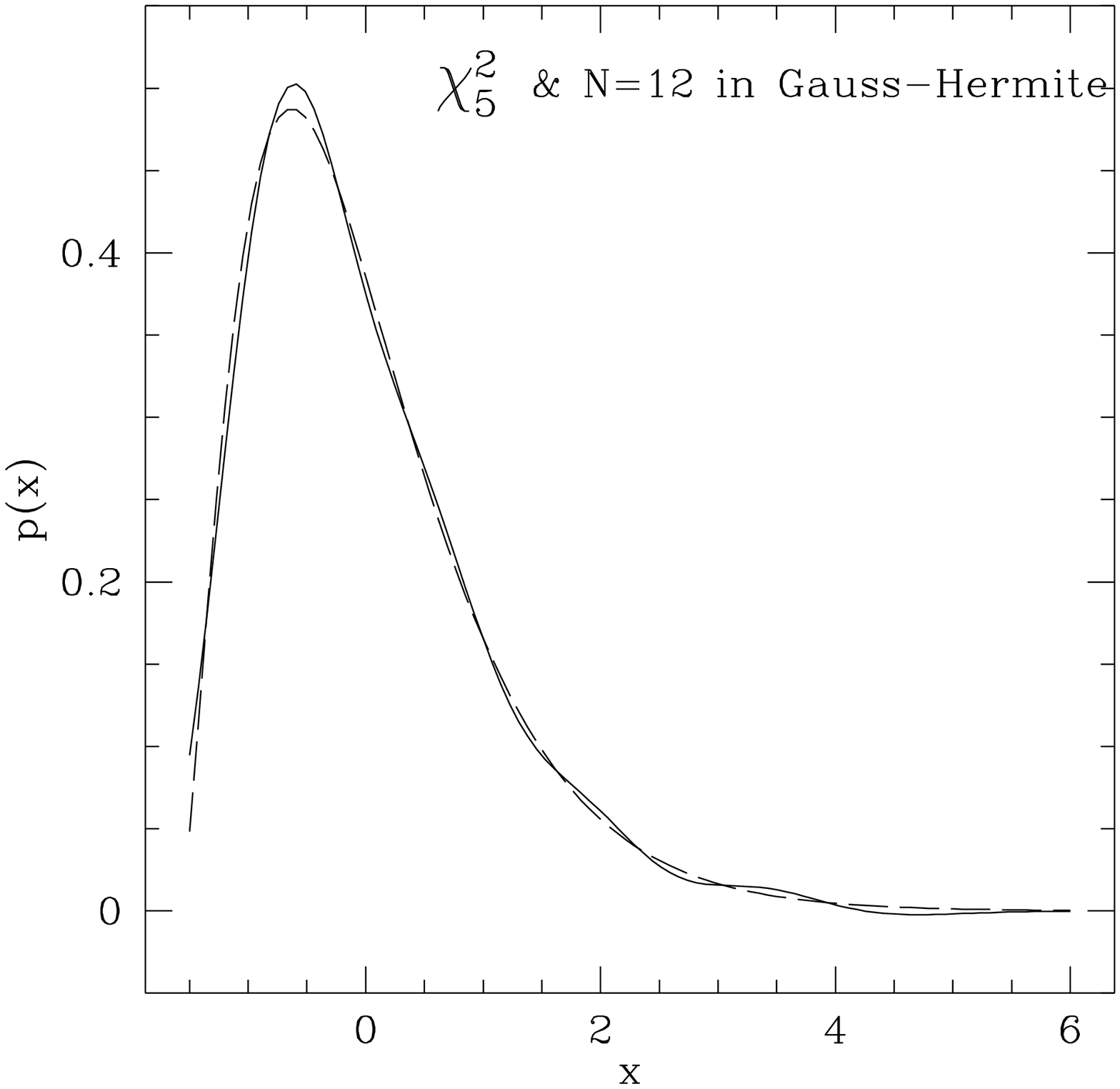}}
  \resizebox{\hsize}{\hsize}{\includegraphics{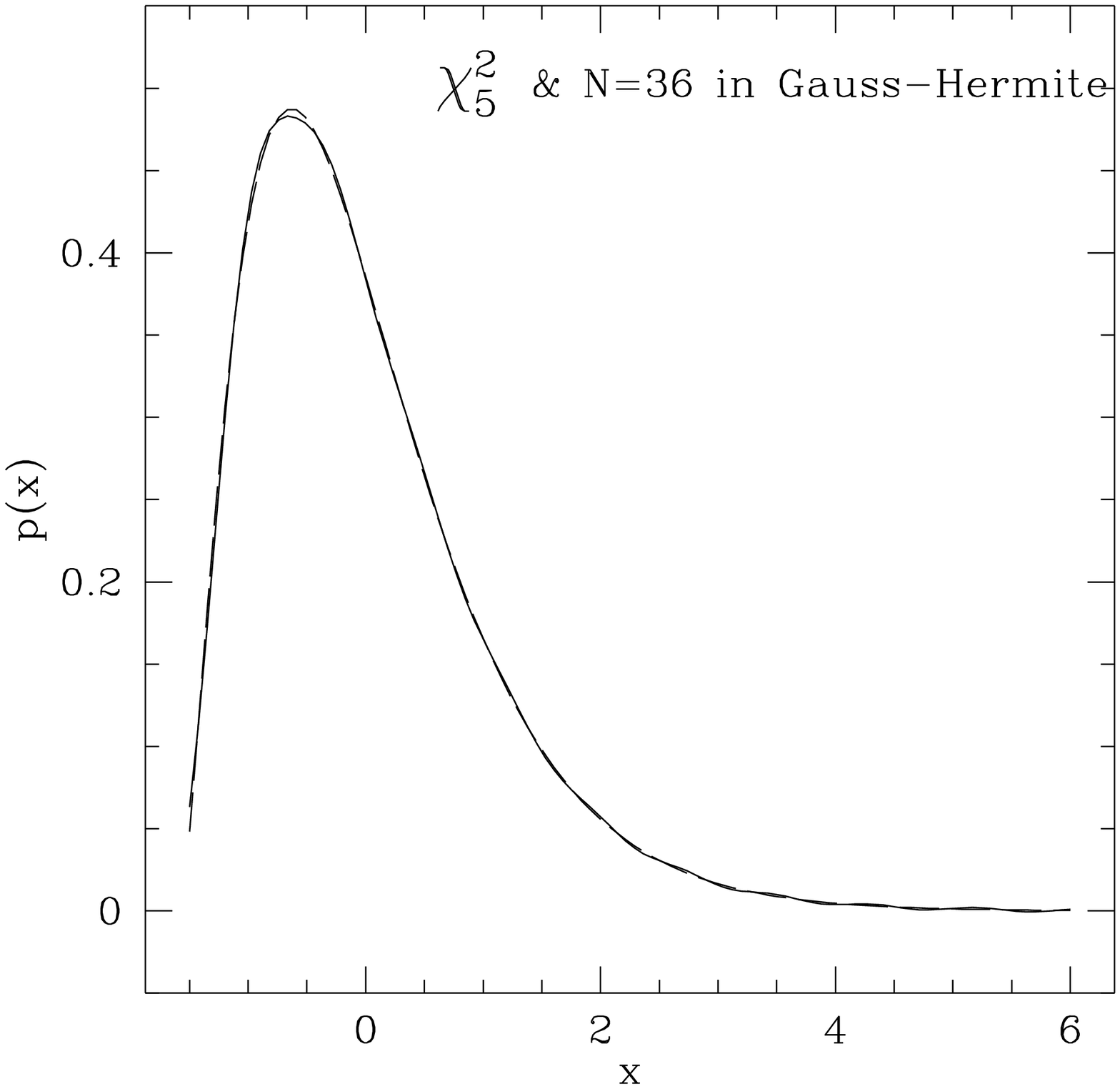}}
\caption{The same as in Fig. \protect\ref{gh5-2-6} but for
         12 and 36 terms in the  Gauss-Hermite expansion}
\label{gh5-12-36}
\end{figure}

The Fourier expansion of $p(x)/Z(x)$ in another set of Hermite polynomials
$H_n(x)$ (\ref{herm}) (not in  Chebyshev-Hermite polynomials $H\!e_n$
(\ref{chherm}),
as for the Gram-Charlier series) is sometimes used:
 \begin{equation}
   p(x) \sim \sum_{n=0}^{\infty} a_nH_n(x)Z(x) \; ,
 \label{gherm}
 \end{equation}
with
 \begin{equation}
   a_n = {\sqrt{\pi}\over 2^{n-1} n!}
         \int_{-\infty}^\infty Z(t) p(t)H_n(t) dt \; .
 \label{ghcoeff}
 \end{equation}
This series is often called the Gauss-Hermite expansion
(see e.g. its application to spectral lines of galaxies
in van der Marel \& Franx \cite{marfra}).
Examples in Figs.~\ref{gh5-2-6} and \ref{gh5-12-36} show its better
convergence.

Van der Marel \& Franx (\cite{marfra})
use a theorem due to Myller-Lebedeff on the convergence of
the Gauss-Hermite expansion: it converges when
$x^3p(x) \rightarrow 0$ for any $p(x)$ with finite and continuous
second derivative. Actually, the conditions sufficient for the
convergence are better: if $p(t)$ obeys the Lipschitz condition
 \begin{equation}
   |p(t)-p(x)| \leq M|t-x|^\alpha, \quad M=\mbox{const},
                                 \quad 0<\alpha\leq 1 \; ,
 \label{lipsch}
 \end{equation}
in a vicinity of $x$ and
 \begin{equation}
    \int_{-\infty}^\infty|p(t)|(1+|t|^{3/2})dt < \infty \; ,
 \label{convergh}
 \end{equation}
then the Gauss-Hermite series converges to $p(x)$ at $x$ (see
e.g. Suetin \cite{suetin}).
These weaker conditions imply that the class of PDFs with
convergent Gauss-Hermite expansions  is much wider than suggested
by the  Myller-Lebedeff theorem cited in 
van der Marel \& Franx \cite{marfra}). But
the simple relation between  the coefficients in the expansion and the
moments of the PDF typical for the Gram-Charlier series is  now lost
[cf. Eq. (\ref{grcoeff}), (\ref{chhermexpl}) and  (\ref{ghcoeff})].
It might be not important in many practical applications when the
moments cannot be accurately determined from observations, but it is
very important for a theoretical work based on the analysis of the moments.

Our Figs.~\ref{gh5-2-6} and \ref{gh5-12-36} show that the 
$\chi^2$ PDF is well-suited for approximation by a  Gauss-Hermite
series. However, one should be
cautious about the accuracy of the computation of the Fourier
coefficients for higher order terms. We have not encountered the
problem of numerical errors in our Gram-Charlier example, since there all
coefficients can be calculated analytically. 
Yet in general care must be taken of the computational accuracy 
to avoid  spurious
numerical divergence of a series which  converges theoretically.

\begin{figure}
  \resizebox{\hsize}{\hsize}{\includegraphics{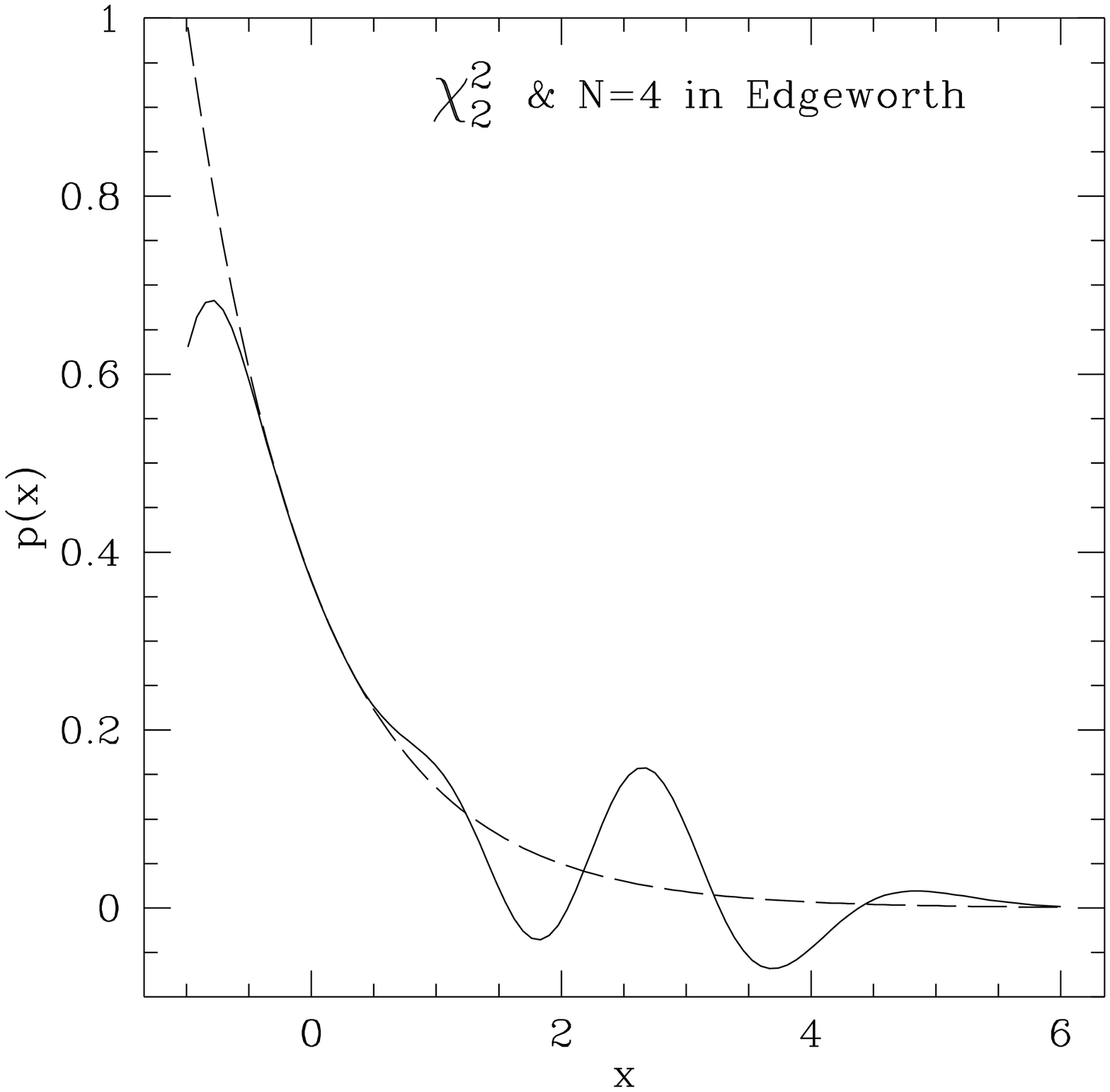}}
  \resizebox{\hsize}{\hsize}{\includegraphics{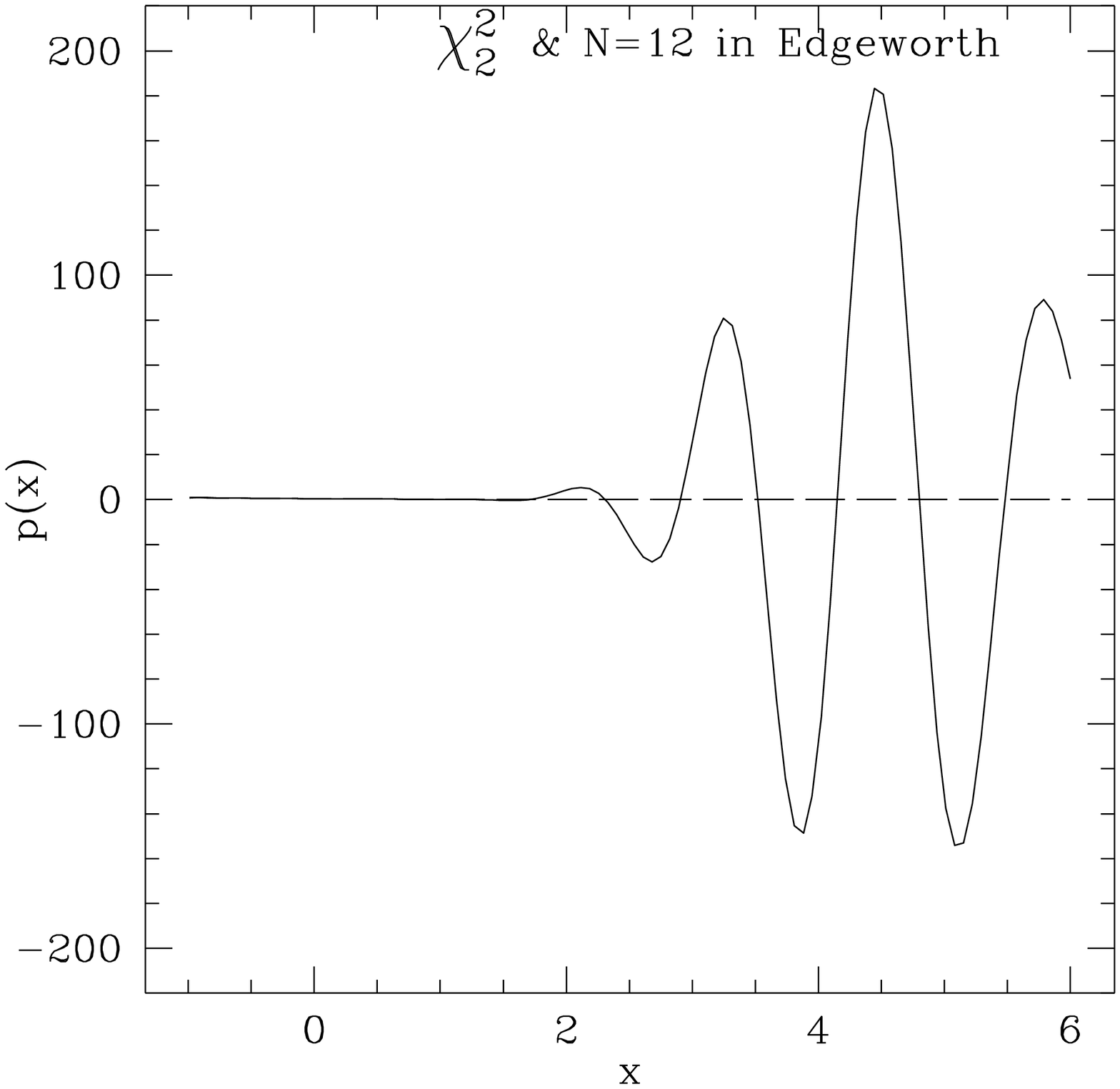}}
\caption{The normalized $\chi^2$ PDF
(\protect\ref{pxden}) for $\nu=2$ (dashed line), and
  its Edgeworth-Petrov approximations
  with 4 and 12 terms in the expansion (solid line).}
\label{ed2-4-12}
\end{figure}

\begin{figure}
  \resizebox{\hsize}{\hsize}{\includegraphics{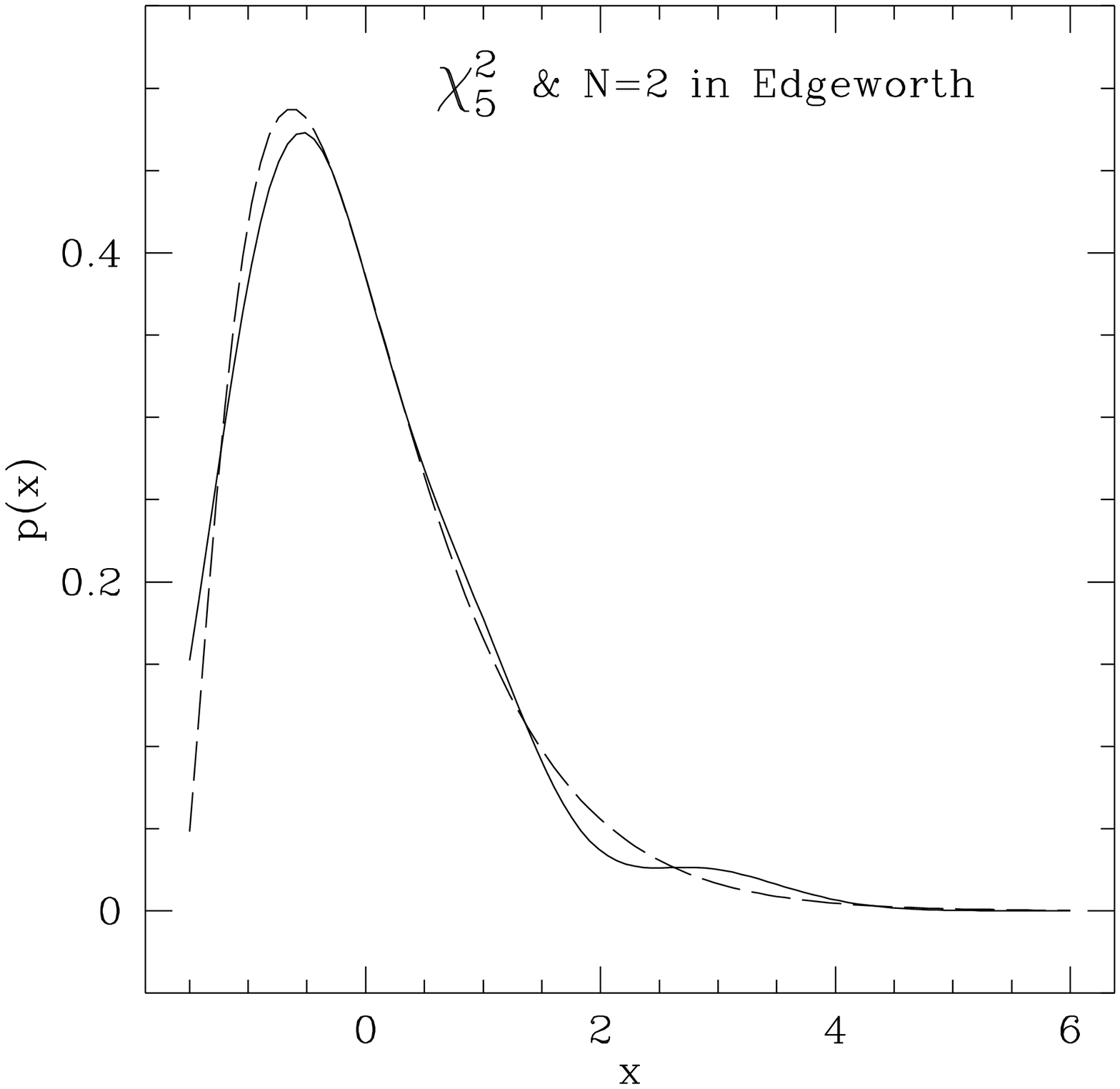}}
  \resizebox{\hsize}{\hsize}{\includegraphics{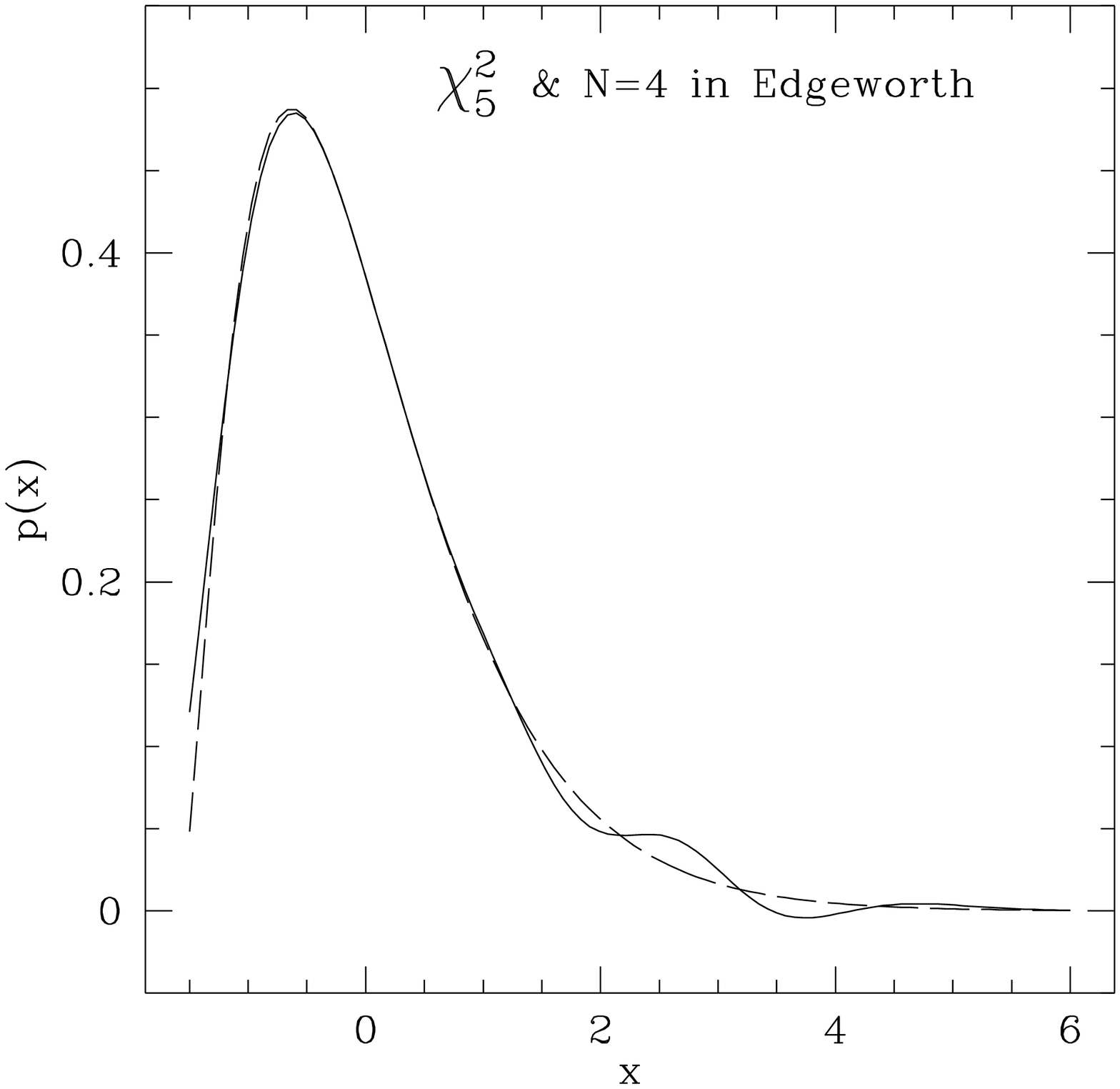}}
\caption{The normalized $\chi^2$ PDF
(\protect\ref{pxden}) for $\nu=5$ (dashed line), and
  its Edgeworth-Petrov approximations
  with 2 and 4 terms in the expansion (solid)}
\label{ed5-2-4}
\end{figure}

\section{The Edgeworth asymptotic expansion}
\label{petr}

A random variable
$X$ can be normalized to unit variance by dividing by its 
standard deviation $\sigma$. The
Edgeworth expansion is a true {\em asymptotic} expansion of the
PDF of the normalized variable $X/\sigma$ in
powers of the parameter $\sigma$, whereas the Gram-Charlier series is
not.
This difference between the Gram-Charlier series and the
Edgeworth expansion was pointed out by Juszkiewicz et al.(\cite{romanetal}),
and in independent work  by Bernardeau \& Kofman
(\cite{bernkof}), followed by Amendola (\cite{amend}) and Colombi 
(\cite{colom}).

 Juszkiewicz et al. (\cite{romanetal}) and Bernardeau \& Kofman 
(\cite{bernkof})  use 2 or 3 terms of the Edgeworth
expansion derived e.g. in Cramer (\cite{cramer}). We note
that the full explicit expansion for arbitrary order  $s$ was obtained
already in 1962 by Petrov (\cite{petrov}), see  also Petrov
(\cite{petr1}, \cite{petr2}).

Petrov derived a powerful generalization of
the Edgeworth expansion for a sum of random variables. In this
section we
give a simplified derivation of the Edgeworth series, following 
Petrov  (\cite{petr1}).
This derivation, for an {\em arbitrary} order, is somewhat 
simpler than the derivation given for example by Bernardeau \& Kofman
(\cite{bernkof}) for the {\em third} order only of
the Edgeworth expansion.

The characteristic function $\Phi(t)$ of a random variable $X$ is
the expectation ${\sf E}\exp(itX)$ as a function of  $t$,
 \begin{equation}
     \Phi(t) \equiv  \int_{-\infty}^{\infty} e^{itx} dF(x) \; ,
 \label{char}
 \end{equation}
that is the Fourier transform of $p(x)$ if the probability density
$p(x)={dF(x)/dx}$ exists. The definition (\ref{char}) implies
that if the moment $\alpha_k$ (\ref{moment}) of $X$ exists,
 \begin{equation}
   \Phi^{(k)}(0)=i^k\alpha_k \; .
 \label{phialph}
 \end{equation}
Hence the Taylor series for $\Phi(t)$ is given by
 \begin{equation}
     \Phi(t) \sim  1 +
        \sum_{k=1}^{\infty} {\alpha_k \over k!} (it)^k\; .
 \label{tayphi}
 \end{equation}
A similar series for $\ln \Phi(t)$,
 \begin{equation}
    \ln \Phi(t) \sim
        \sum_{n=1}^{\infty} {\kappa_n \over n!} (it)^n\; ,
 \label{taylnphi}
 \end{equation}
involves cumulants (semi-invariants) $\kappa_n$ defined by
 \begin{equation}
      \kappa_n \equiv {1\over i^n}
      \Bigr[{ d^n \over dt^n} \ln\Phi(t) \Bigl]_{t=0} \; .
 \label{cumul}
 \end{equation}

In   \ref{Lemma}  we prove  a fundamental lemma of calculus for the
$n-$th derivative of a composite function $f\!\circ\! g(x) \equiv f(g(x))$,
which reads
 \begin{eqnarray}
 \lefteqn{{d^n \over dx^n} f(g(x)) =} \nonumber \\ 
 \lefteqn{\qquad   n! \sum_{\{k_m\}}f^{(r)}(y)|_{y=g(x)}
         \prod_{m=1}^n {1 \over k_m!}
         \left({1 \over m!} g^{(m)}(x)\right)^{k_m} \; ,}
 \label{lemma}
 \end{eqnarray}
where  $r=k_1+k_2+ ... +k_n$ and
the set $\{k_m\}$ consists of all non-negative integer solutions of
the Diophantine equation
 \begin{equation}
 k_1+2k_2+ ... +n k_n=n \; .
 \label{dioph}
 \end{equation}

Using (\ref{lemma}) we derive
a useful relation (Petrov \cite{petr2})
between the cumulants $\kappa_n$ and the moments $\alpha_k$ of a
PDF  in   \ref{Lemmapl},
 \begin{equation}
      \kappa_n =
      n! \sum_{\{k_m\}}(-1)^{r-1}(r-1)!
         \prod_{m=1}^n {1 \over k_m!}
         \left({\alpha_m \over m!} \right)^{k_m} \; .
 \label{cumalph}
 \end{equation}
Here summation extends over all non-negative integers $\{k_m\}$
satisfying (\ref{dioph})  and $r=k_1+k_2+ ... +k_n$.
We describe a simple algorithm for obtaining all solutions of
Eq. (\ref{dioph}) in \ref{algor}.

Now we are ready to begin with the derivation of the Edgeworth
expansion. Consider
a random variable $X$ with ${\sf E}X=0$ (this can  always be achieved
by an appropriate choice of origin), and let $X$ have 
dispersion $\sigma^2$.
If $X$ has the characteristic function $\Phi(t)$, then the
normalized random variable $X/\sigma$ has the characteristic function
$\varphi(t)=\Phi(t/\sigma)$. Therefore we have  from Eqs.
(\ref{taylnphi}) and  (\ref{cumul}) that 
\begin{equation}
    \ln \varphi(t)=\ln \Phi(t/\sigma) \sim
        \sum_{n=2}^{\infty} {\kappa_n \over \sigma^n n!} (it)^n\;  .
 \label{taylnf}
 \end{equation}
Here the sum starts at $n=2$ because ${\sf E}X=0$.  Moreover, since 
$\kappa_2=\sigma^2$ (see Eq.~(\ref{cumalph}))
we obtain
 \begin{equation}
     \varphi(t) \sim e^{-t^2/2} \exp \left\{
        \sum_{n=3}^{\infty} { S_n \sigma^{n-2} \over n!} (it)^n
                           \right\} \; ,
 \label{taylphi}
 \end{equation}
with
 \begin{equation}
      S_n \equiv  \kappa_n / \sigma^{2n-2} \; .
 \label{normcumul}
 \end{equation}                                
Let us write the exponential function in (\ref{taylphi}) as a
formal series in powers of $\sigma$,
 \begin{equation}
      \exp \left\{
        \sum_{r=1}^{\infty} { S_{r+2} \sigma^r \over (r+2)!}
        (it)^{r+2} \right\} \sim 1 + \sum_{s=1}^\infty
        {\cal P}_s (it)\sigma^s \; ,
 \label{powser}
 \end{equation}
where the coefficient of the power $s$ is a function
${\cal P}_s(it)$. Now, using 
$g(x) \equiv \sum_{r=1}^{\infty}\{S_{r+2}(it)^{r+2} x^r /(r+2)!\}$
and $f\equiv \exp$ in (\ref{lemma}), we find that
 \begin{eqnarray}
 \lefteqn{   {\cal P}_s(it) \equiv { 1 \over s!}
   {d^s \over dx^s} f(g(x))|_{x=0} }  \nonumber \\
 \lefteqn{\qquad  = \sum_{\{k_m\}}\prod_{m=1}^s {1 \over k_m!}
         \left({S_{m+2} (it)^{m+2} \over (m+2)!}\right)^{k_m} \; ,}
 \label{pn}
 \end{eqnarray}
where the summation extends again over all non-negative integers $\{k_m\}$
satisfying (\ref{dioph}). Thus the function ${\cal P}_s$ is just
a polynomial.

Suppose that the probability density $p(x)$ of a random variable $X$
exists. Then the PDF for $X/\sigma$  is
$q(x) \equiv \sigma p(\sigma x)$, and
it is the inverse Fourier transform of the characteristic function 
$\varphi$:
 \begin{equation}
   q(x) =
        {1\over 2\pi} \int_{-\infty}^{\infty} e^{-itx}\varphi(t) dt \; .
 \label{invchar}
 \end{equation}
If $\Phi(t)$ is the Fourier transform of a function $p(x)$,
then $(-it)^n\Phi(t)$  is the transform of the $n$-th
derivative of $p(x)$,
 \begin{equation}
    {d^n \over dx^n} p(x) = {1\over 2\pi}
          \int_{-\infty}^{\infty} e^{-itx}(-it)^n \Phi(t) dt \; .
 \label{derfour}
 \end{equation}
The Fourier transform of the Gaussi\-an
distribution $Z(x)$ in (\ref{ggaus}) is $\exp(-t^2/2)$ ,
see e.g. Bateman \& Erd\'elyi (\cite{baterd}).
Therefore  each $(it)^n$,  multiplied by $\exp(-t^2/2)$ in the
expansion of $\varphi$ (see Eqs.~\ref{taylphi} to 
\ref{pn}),  generates  (according to Eq.~\ref{invchar}) the $n$-th
derivative of $Z(x)$,
 \begin{equation}
   (-1)^n {d^n \over dx^n} Z(x)   =
         \int_{-\infty}^{\infty} e^{-itx}(it)^n \exp(-t^2/2) dt \; ,
 \label{derng}
 \end{equation}
in the corresponding expansion for $q(x)$,
 \begin{eqnarray}
 \lefteqn{  q(x)=  Z(x) +  \sum_{s=1}^\infty \sigma^s }  \nonumber \\
 \lefteqn{      \times \left\{ \sum_{\{k_m\}}\prod_{m=1}^s {1 \over k_m!}
         \left({S_{m+2} (-1)^{m+2} \over (m+2)!}
   {d^{m+2} \over dx^{m+2}} \right)^{k_m} Z(x) \right\} \; .}
 \label{preedgew}
 \end{eqnarray}
Here the set $\{k_m\}$ in the sum consists of all non-negative
integer solutions of the equation
 \begin{equation}
 k_1+2k_2+ ... +s k_s=s \; .
 \label{diophnu}
 \end{equation}
Using (\ref{derg}) and $r=k_1+k_2+ ... +k_s$ we can rewrite
(\ref{preedgew}) in terms of the Chebyshev-Hermite polynomials:

\begin{eqnarray}
 \lefteqn{ q(x)  =   \sigma p(\sigma x)  =   Z(x) 
    \left\{ 1 + \sum_{s=1}^\infty \sigma^s \right. }\nonumber  \\
 \lefteqn{\quad  \times  \left.  \sum_{\{k_m\}} H\!e_{s+2r}(x)  
      \prod_{m=1}^s {1 \over k_m!} \left({S_{m+2} \over (m+2)!}       
                           \right)^{k_m} \right\}  \; .}
 \label{edgew}
\end{eqnarray}

%

This is the Edgeworth expansion for arbitrary order
$s$. See Petrov (\cite{petr1},\cite{petr2}) for a more general form of the
expansion (for non-smooth cumulative distribution functions $F(x)$ and for
a sum of random variables) and for the proof that the series
(\ref{edgew}) is asymptotic (see also the classical references Cram\'er 
\cite{cramer} and Feller \cite{feller}). This means that
if the first $N$ terms are retained in the sum over $s$, then the
difference between $q(x)$ and  the partial sum is of a lower order
than the $N$-th term in the sum (Erd\'elyi \cite{erd}; Evgrafov
\cite{evgr}). Convergence
plays no role in the definition of the asymptotic series.

Strictly speaking, Petrov  (\cite{petr1}) 
proves the asymptotic theorems for
sums of $\nu$ independent random variables only when $\sigma \sim 1/\nu^{1/2}$,
and not for any $\sigma$ , which we used in our derivation. But in all
practical applications where  nearly Gaussian PDFs occur (and in all
applications that we consider in the present work), those PDFs basically 
 {\em are} the sums of random variables, and the proofs of the asymptotic 
theorems are relevant. In the next section we show how the theory
works in practice.

\begin{figure}
  \resizebox{\hsize}{\hsize}{\includegraphics{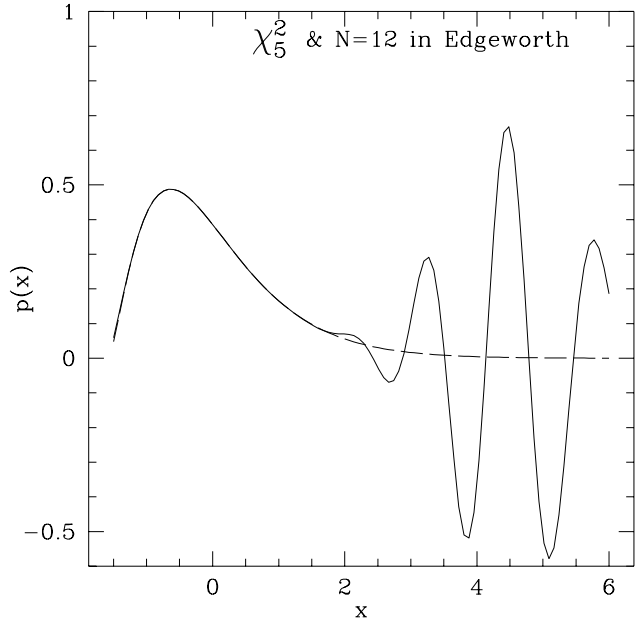}}
\caption{The normalized $\chi^2$ PDF
(\protect\ref{pxden}) for $\nu=5$ (dashed line) and
  its Edgeworth-Petrov approximations
  with  12 terms in the expansion (solid)}
\label{ed5-12}
\end{figure}

\begin{figure}
  \resizebox{\hsize}{\hsize}{\includegraphics{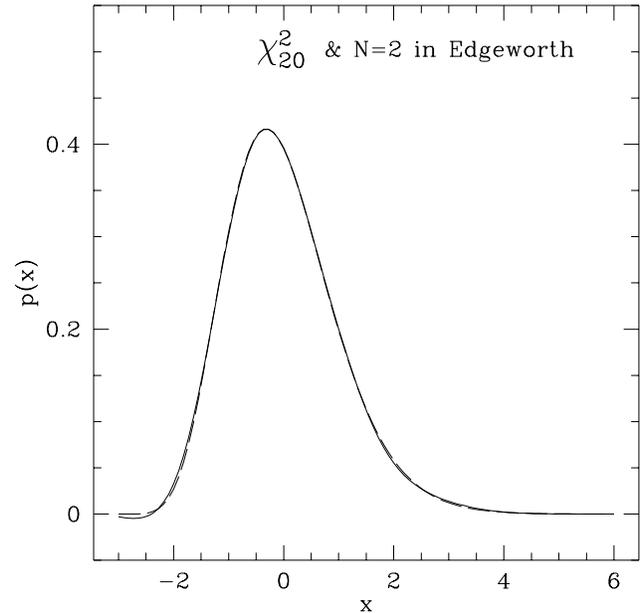}}
  \resizebox{\hsize}{\hsize}{\includegraphics{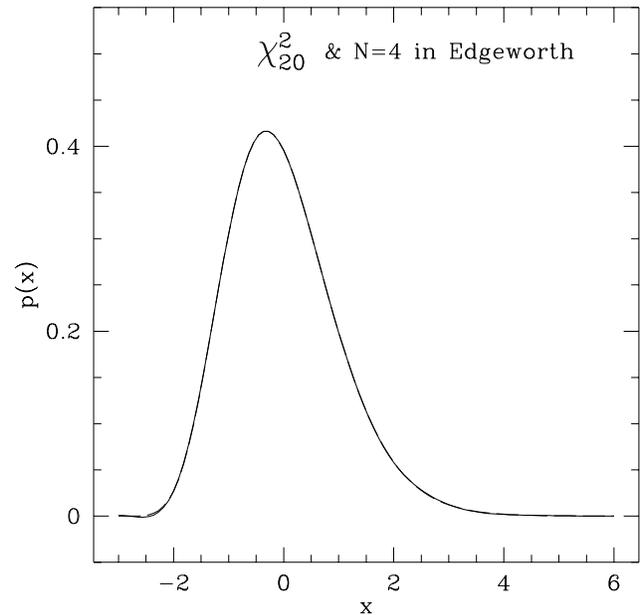}}
\caption{The normalized $\chi^2$ PDF
(\protect\ref{pxden}) for $\nu=20$ (dashed line), and
  its Edgeworth-Petrov approximations
  with 2 and 4 terms in the expansion (solid)}
\label{ed20-2-4}
\end{figure}

Figures \ref{ed2-4-12} -- \ref{ed20-2-4} show 
some examples of the Edgeworth expansion for the
$\chi^2$ distribution. It is clear that for strongly non-Gaussian
cases, like $\chi^2_\nu$ for $\nu=2$, it has a very small domain of applicability
in practical cases since  it diverges like the Gram-Charlier series 
for a large number of terms (Fig. \ref{ed2-4-12}).
But already
in this case one can check that the order of the last term retained
gives the order of the error correctly, and one can truncate the 
expansion when the last term becomes unacceptably large. 
For nearly Gaussian distributions the situation is much better: compare 
the cases for $\nu=5$ and $\nu=20$ in Figs. \ref{ed5-2-4}, \ref{ed5-12} and 
\ref{ed20-2-4}.

\begin{figure}
  \resizebox{\hsize}{\hsize}{\includegraphics{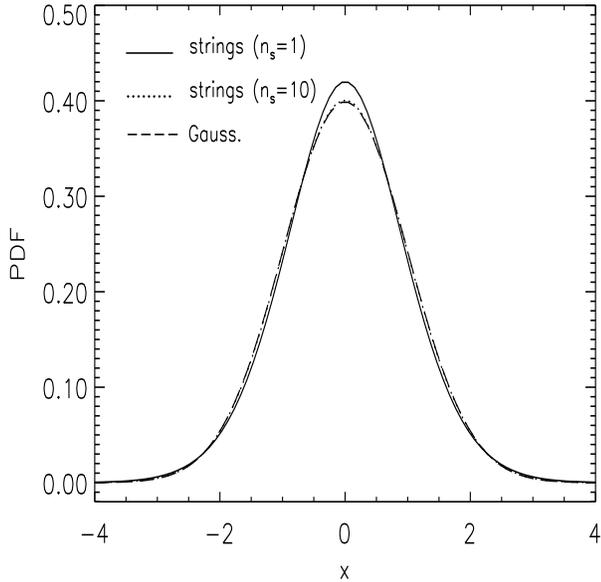}}
\caption{Edgeworth expansion up to 10th order for the PDF of peculiar 
         velocities from cosmic strings, within an analytical  model 
         for the string network, for two values of the 
         number of strings per Hubble volume (Moessner \cite{moessn}).}
\label{cspdf}
\end{figure}

\begin{figure}
  \resizebox{\hsize}{\hsize}{\includegraphics{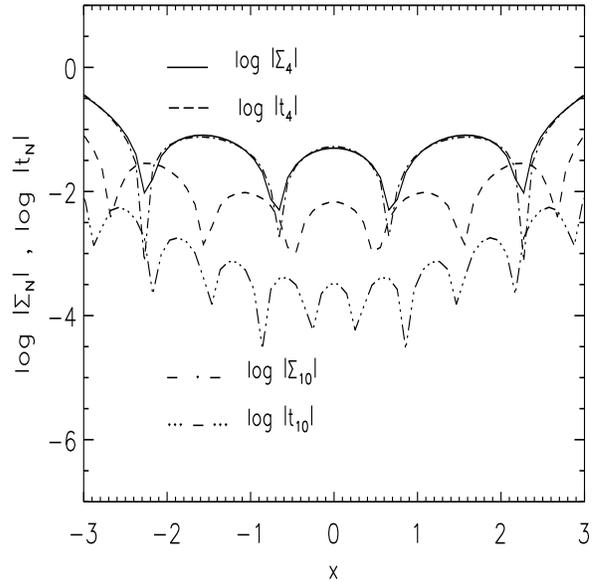}}
\caption{Relative deviation $\Sigma_N$ from a normal distribution of
the Edgeworth expansion up to $N$th order of the PDF of peculiar
velocities from cosmic strings, for $n_s=1$. Also shown is the error $t_N$
of this deviation associated with the expansion.}
\label{csdev}
\end{figure}

\begin{figure}
  \resizebox{\hsize}{\hsize}{\includegraphics{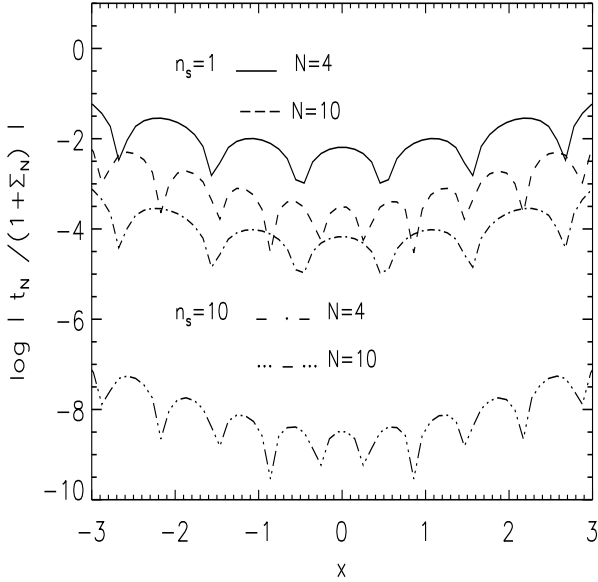}}
\caption{Relative error $t_N/(1+\Sigma_N)$ in the Edgeworth expansion
of $q(x)/Z(x)$ for the PDF of peculiar velocities from cosmic strings,
for two values of $N$ and $n_s$.}
\label{csrel}
\end{figure}

\section{Peculiar velocities from cosmic strings}
\label{CS}

As an example we consider the probability distribution of
peculiar velocities within the cosmic string model of structure 
formation. 
Cosmic strings are one-dimensional topological defects possibly formed
in a phase transition in the early Universe (Brandenberger
\cite{brand}; Hindmarsh \& Kibble \cite{hindm}; Vilenkin \& Shellard 
\cite{vilenk}). After the time of 
formation, the string network quickly evolves to a scaling solution 
with a constant number $n_s$ of strings passing through a  Hubble volume 
in one expansion time.

Since individual topological defects give rise to velocity 
perturbations of the dark matter through which they move up to a
distance of a Hubble radius, one expects a non-Gaussian result, 
in contrast to inflationary theories which predict a Gaussian
PDF. Therefore departures from a normal distribution may be 
a way to distinguish between the two main classes of theories of
structure formation, inflation and topological defects. However, since
many strings present between the time of equal matter and radiation
contribute to the perturbations, a nearly Gaussian PDF can result due to 
the central limit theorem. 

In order to estimate the deviation of the PDF from the normal distribution
we use Petrov's formula  (\ref{edgew}) for the Edgeworth series
in this section.
We calculate  the cumulants up to 12th order in the analytical  model for 
the cosmic string network presented in Moessner (\cite{moessn}), 
where the cumulants are given up to 8th order.  

For simplicity, let us write Eq.~\ref{edgew} schematically as 
\begin{equation}
q(x)=Z(x) \{ 1+ \sum_{s=1}^\infty t_{s} \}  \; ,
\label{simple}
\end{equation}
i.e. denote the $s$th term in the sum by $t_{s}$. Let us further 
denote the sum  up to $s=N$ by $\Sigma_N \equiv \sum_{s=1}^N t_{s}$. 
In Fig.~\ref{cspdf} we show the Edgeworth expansion of the PDF of
peculiar velocities up to 10th order, for the case of 
$n_s$=1 string per Hubble volume. 
Expanding up to $N$th order, the relative deviation of the PDF from a normal 
distribution is given by,
\begin{equation}
\frac{q(x)}{Z(x)}-1 = \Sigma_N  \; .
\end{equation}
It is only significant if the error of the asymptotic expansion,
which is of the order of the last term included, is smaller
than this deviation. 
In Fig.~\ref{csdev} we show the relative deviation $\Sigma_N$ and the 
error $t_N$ associated with it. 
The wiggles or cusps in the graphs appear at zeros of $\Sigma_N$ and
$t_N$ which are both oscillating, changing their sign, and we plot
the logarithms of the absolute values. So only the maxima of the curves give
a true indication  of the deviations and errors.
For $N=10$ the error is clearly below
the deviation and thus significant. For $N=4$, the error is still 
below the deviation for most values of $x$, but it is not as clear,
especially since the error is not exactly equal to $t_N$ but of that
order only.
In Fig.~\ref{csrel} we show the relative error
$t_N/(1+\Sigma_N)$ in the expansion of $q(x)/Z(x)$. It
is smaller for an expansion to higher order for a given number  of
strings per Hubble volume. The error decreases more strongly with $N$ 
for larger $n_s$. 
The relative error is also smaller, at fixed $N$, for a  larger $n_s$,
in which case the PDF is closer to a normal distribution. 

It is interesting to compare these results with the general theory.
Petrov  (\cite{petr1}) proves that under certain conditions (which
are fulfilled in most physically important cases)
\begin{equation}
q(x) = Z(x)\{ 1+ \sum_{s=1}^N t_{s} \} +o(\sigma^N)  \; 
\label{asymp}
\end{equation}
uniformly for $-\infty < x < +\infty$, when $\sigma \sim 1/\nu^{1/2}$
and the PDF $q(x)$ is for a sum of $\nu$ random variables. One
should remember that each $t_s$ is of the order of $\sigma^s$ in (\ref{edgew}).
In our case $\nu$ is just $n_s$, so the error of the truncated
Edgeworth series scales as $\sim1/n_s^{N/2}$. From Fig.~\ref{csrel} we can
 see that the error for the case of $n_s=10$ strings is indeed
$N/2$ orders less than for the case of $n_s=1$ string, i.e. 2 orders
for $N=4$ and 5 orders for $N=10$ terms in the expansion. This is an 
illustration of  the  theory developed by
Petrov  (\cite{petr1},\cite{petr2}).

\section{Conclusions}
\label{concl}

We have shown that the Gram-Charlier series has a  limited
domain of applicability for nearly normal distributions because of
its rather poor convergence properties. The Gauss-Hermite expansion 
can give good results in problems like fitting profiles of spectral
lines of galaxies, supernovae, or ordinary stars. In advanced 
 calculations of stellar atmospheres (e.g. Hauschildt et al. \cite{hba};
Hubeny \&  Lanz \cite{hublan}) the profiles of thousands or even millions 
of lines
must be integrated for up to hundreds of Doppler widths, and the 
Gauss-Hermite expansion can perhaps be useful for saving information
of the line profiles in an economical way. But since it has no intrinsic
measure of accuracy, the number of
terms needed  in the expansion must be examined carefully for each individual 
problem.

For situations where the estimate of a deviation of a PDF from a Gaussian 
one is needed, the asymptotic Edgeworth expansion is indispensable,
and for high order moments the form of this expansion found
by Petrov is necessary. We found a workable 
algorithm for Petrov's formula of the Edgeworth
expansion and applied it to  several examples.
The source codes used in this work are available on request from the authors.

\begin{acknowledgements}

We are grateful to Ya.M.Kazhdan for valuable advice on references, 
to N.N.Pavlyuk 
for assistance, and to the referee for useful comments. The work of SB 
is supported in part by INTAS grant
"Thermonuclear Supernovae" and by a grant from the Research Center
for the Early Universe, University of Tokyo, and he cordially thanks K.Nomoto,
as well as W.Hillebrandt, MPA, Garching, for their hospitality.

This research has made use of NASA's Astrophysics Data System Abstract Service.

\end{acknowledgements}

\vspace{5mm}

\appendix
\section*{Appendices}
\section{Lemma}
\label{Lemma}
In Sect. \ref{petr}, the relation (\ref{lemma}) for the $n-$th derivative of a
composite function $f\!\circ\! g(x) \equiv f(g(x))$ is used 
for the derivation of the Edgeworth asymptotic expansion.
Here a simplified derivation of Eq.(\ref{lemma}) is given.
Petrov (\cite{petr1},\cite{petr2}) suggests a proof 
by induction. We note that this derivation is more transparent
if one simply considers the Taylor expansion for $f\!\circ\! g$ expressed
in terms of the Taylor expansions of $f$ and $g$ -- see Bourbaki (\cite{bourb}).
We have
 \begin{equation}
   f(y) = f(y_0) + {f'\over 1!}\Delta y +
                   {f''\over 2!}\Delta^2 y+ \cdots
              {f^{(n)} \over n!}\Delta^n y+ \cdots \; 
 \label{tayf}
 \end{equation}
and
 \begin{equation}
   g(x) = g(x_0) + {g'\over 1!}\Delta x +
                   {g''\over 2!}\Delta^2 x+ \cdots
              {g^{(m)} \over m!}\Delta^m x+ \cdots \; .
 \label{tayg}
 \end{equation}
Truncating the expansions at some $n$ and $m$ we find that
 \begin{eqnarray}
 \lefteqn{  f\!\circ\! g(x) = f(g(x_0)) } \nonumber \\
 \lefteqn{\quad   + {f'\over 1!}    \left({g'\over 1!}\Delta x
          +  {g''\over 2!}\Delta^2 x+ \cdots
              {g^{(m)} \over m!}\Delta^m x+ \cdots \right)} \nonumber \\
 \lefteqn{\quad  +\cdots} \\
 \lefteqn{ \quad +    {f^{(n)} \over n!}
   \left({g'\over 1!}\Delta x +
                   {g''\over 2!}\Delta^2 x+ \cdots
              {g^{(m)} \over m!}\Delta^m x+ \cdots \right)^n \; .} \nonumber
 \label{tayfg}
 \end{eqnarray}
On the other hand, we can write down the Taylor series for the
composite function,
 \begin{eqnarray}
 \lefteqn{   f\!\circ\! g(x) = f\!\circ\! g(x_0)}  \nonumber \\
 \lefteqn{\quad  + {(f\!\circ\! g)'\over 1!}\Delta x +
                   {(f\!\circ\! g)''\over 2!}\Delta^2 x+ \cdots
              {(f\!\circ\! g)^{(s)} \over s!}\Delta^s x+ \cdots  }
 \label{taycompfg}
 \end{eqnarray}
Now using the polynomial theorem,
 \begin{eqnarray}
  \lefteqn{   (x_1+x_2+\cdots+x_m)^r } \nonumber \\
 \lefteqn{\quad  =  \sum_{\{k_m\}}
  { r! \over k_1! k_2! \cdots k_m!} x_1^{k_1}x_2^{k_2}\cdots x_m^{k_m} }
                                        \nonumber \\
 \lefteqn{\quad = \sum_{\{k_m\}} r!\prod_{s=1}^m {x_s^{k_s} \over k_s!} 
     \; , }
 \label{polynth}
 \end{eqnarray}
where summation extends over all sets of non-negative integers $\{k_m\}$
satisfying  $r=k_1+k_2+ ... +k_s$,
and comparing the terms $\Delta^{s}x$ with equal $s$ in
(\ref{tayfg}) and (\ref{taycompfg}),  we obtain
 \begin{eqnarray}
  \lefteqn{ {d^s \over dx^s} f(g(x)) = }  \nonumber \\
  \lefteqn{ s! \sum_{\{k_m\}}f^{(r)}(y)|_{y=g(x)}
	 \prod_{m=1}^s {1 \over k_m!}
         \left({1 \over m!} g^{(m)}(x)\right)^{k_m} \; .}
 \label{lemma1}
 \end{eqnarray}
This is the relation (\ref{lemma}) which we sought.
 Here the set $\{k_m\}$ consists of non-negative solutions
of the Diophantine equation
 \begin{equation}
   k_1+2k_2+ ... +s k_s=s \; ,
 \label{dioph1}
 \end{equation}
and  $r=k_1+k_2+ ... +k_s$.

\section{Applications of the Lemma}
\label{Lemmapl}

If we apply lemma (\ref{lemma})
to Chebyshev-Hermite polynomials in (\ref{chherm}), we 
$g(x)=-x^2/2$ for $f=\exp$, so that only the terms with $m=1$ and
$m=2$ are non-zero in the product in (\ref{lemma}), and we only need
non-negative integers $\{k_1, k_2\}$ as the solutions for
$k_1+2k_2=n$. Thus for each $k\equiv k_2$ running from $0$ to
$[n/2]$ ({\em entier} of $n/2$) we have $k_1=n-2k$ and
$r=n-k$. Finally, we have  from (\ref{lemma}) that 
 \begin{eqnarray}
  \lefteqn{ {d^n \over dx^n}\exp(-x^2/2) = 
    n! \sum_{k=0}^{[n/2]} e^{-x^2/2} 
         \ {1 \over (n-2k)!} }   \nonumber \\
  \lefteqn{  \qquad \qquad  \times  \left({1 \over 1!} (-x)\right)^{n-2k}
         \ {1 \over k!}
         \left({1 \over 2!} (-1)\right)^{k} \; , }
 \label{derexp}
 \end{eqnarray}
and the explicit expression (\ref{chhermexpl}) follows immediately
from Rodrigues' formula  (\ref{chherm}).

Among other consequences of (\ref{lemma}) is the
relation (\ref{cumalph}) between cumulants $\kappa_n$ and moments
$\alpha_k$ of a
PDF. From the definition (\ref{cumul}) we obtain this relation by simply
applying
(\ref{lemma})  to the case of $f\equiv \ln$ and $g\equiv\Phi$. Since
$f^{(r)}(y)|_{y=g(t)}=(-1)^{r-1}(r-1)!/\Phi^r|_{t=0}=(-1)^{r-1}(r-1)!$, 
we find that
 \begin{eqnarray}
   \lefteqn{ \kappa_n = {1\over i^n}
  {d^n \over dt^n} \ln\Phi|_{t=0} = }\nonumber \\
  \lefteqn{ {n!\over i^n}
        \sum_{\{k_m\}}(-1)^{r-1}(r-1)!
         \prod_{m=1}^n {1 \over k_m!}
         \left({1 \over m!} \Phi^{(m)}|_{t=0}\right)^{k_m} \; .}
 \label{cumphi}
 \end{eqnarray}
Thus, from (\ref{phialph}),
 \begin{equation}
      \kappa_n =
      {n!\over i^n}  \sum_{\{k_m\}}(-1)^{r-1}(r-1)!
         \prod_{m=1}^n {i^{mk_m} \over k_m!}
         \left({\alpha_m \over m!} \right)^{k_m} \; ,
 \label{cummom}
 \end{equation}
which is equivalent to (\ref{cumalph}).
Here the sum extends over all non-negative integers $\{k_m\}$
satisfying (\ref{dioph})  and $r=k_1+k_2+ ... +k_n$ .

\section{Algorithm for computing indices $k_m$}
\label{algor}

To find all the solutions of Eq. (\ref{dioph}) 
it is desirable to order  all
sets of non-negative integers $\{k_m\}$ satisfying it.  
We first rewrite (\ref{dioph}) as  
 \begin{equation}
 nk_n + ... +2k_2+  k_1=n \; .
 \label{diophrev}
 \end{equation}
It is natural to
establish the ordering of the sets $S_i\equiv\{k_m\}$ satisfying 
Eq. (\ref{dioph}) 
according to the order of numbers in their decimal representation. 
For $n=3$, say, we have 3 non-negative solutions 
$$S_1=\{k_3=0,k_2=0,k_1=3\}$$
$$S_2=\{k_3=0,k_2=1,k_1=1\}$$
$$S_3=\{k_3=1,k_2=0,k_1=0\}$$
which can  simply be written as
$$S_1=003, \; S_2=011, \; S_3=100 \; ,$$
and we say that $S_1<S_2<S_3$ since $3<11<100$. For  $n\ge 10$,
when the base $10$ is no longer convenient, the sets of solutions
can be ordered according to the  order of the integer numbers  
 \begin{equation}
 k_n r^{n-1} + ... +k_2 r+  k_1  
 \label{korder}
 \end{equation}
for any natural base $r$. Those numbers are not important in themselves,
what matters is thinking about the sets of $\{k_m\}$ as numbers
in an abstract  representation to base $r$.

On entry to our algorithm we have an integer ${\tt n} > 0$, and
on exit we wish to have the number {\tt nsol}  of all solutions
of Eq. (\ref{dioph}) and the solutions themselves. For a set $S_i$ we 
introduce an abstract variable {\tt S}
to describe the array {\tt k(1:n)} of {\tt n}  integer  elements 
ordered as ``quasi-decimal digits''. 

So, in an abstract form the algorithm looks like:

\begin{verbatim}
  --     Statement QS: 
  --             all S <= S(CURRENT) 
  --             are out if and only if 
  --             k(1)+2*k(2)+...+n*k(n)=n 
  <*initiate: make QS true for S=S(INITIAL) *>; 
  _while <*condition:  S ^= S(FINAL) *> _do 
    <*nextset: find next  S  
                        keeping QS invariant *> 
  _od; -- Here S=S(FINAL) and QS=.TRUE., i.e.
       -- all needed solutions are found
\end{verbatim}
It is easy to {\tt initiate QS}:

\begin{verbatim}
       k(1)=n; 
       mold=1; -- keeps largest m 
               -- for non-zero k(m) 
       _do m=2,n; 
         k(m)=0
       _od; 
       nsol=1; 
\end{verbatim}
The integer {\tt mold}  keeps track of the progress
of the algorithm and allows us to establish the {\tt condition} for 
the continuation of the main loop,
$$
       {\tt mold}\ne{\tt n} \; .
$$
Finally, the core of the algorithm is the node {\tt nextset}
where we work with our sets as with digits, which is like
doing  the addition 
of decimal numbers by hand, column by column from right to left.

\begin{verbatim}
  --   advance S(CURRENT), i.e. consider adding
  --   n-( 2*k(2) +3*k(3) + ...+mold*k(mold) )
  --   to k(1) trying to add 1 to the lowest of
  --   k(2),...,k(mold) possible.
  --   If adding 1 to the lowest k(m) 
  --   makes the sum > n then
  --   put this k(m)=0 and try k(m+1) 
  -- 
   m=1;                                            
   sumcur=n; -- integer sum of current set S       
   _repeat                                         
       sumcur=sumcur-k(m)*m+m+1;                   
       k(m)=0;                                     
       k(m+1)=k(m+1)+1;                            
       m=m+1;                                      
   _until sumcur <= n _or  m>mold;                 
   _if m>mold _then mold=m _fi;                    
   k(1)=n-sumcur;                                  
   nsol=nsol+1;                                    
\end{verbatim}

Here the node  {\tt nextset} ends, and the whole  algorithm 
is finished. We have written it here in a pseudocode which
can be translated mechanically into any machine language. We actually
use a special preprocessor {\tt Trefor} 
which automatically transforms
the text above to standard Fortran (see Weinstein \& Blinnikov
\cite{trefor}, and Bartunov et al. \cite{trfguide}).

For reference, we present the first 8 sets of solutions found 
by this algorithm.
In practice it is easier not to use the table below even for low {\tt n},
but generate all coefficients in the code.

\begin{table}
\caption{Table of solutions to Eq. (\protect\ref{dioph}) for $n=1$ to 8.
Zeros for higher orders are left blank }
\begin{center}
\begin{tabular}{rrrrr}
\hline
n  & 1     &  2     &  3   &    4 \\
\hline
&1     &  2     &  3   &    4     \\
&      & 10     & 11   &   12     \\
&      &        &100   &   20     \\
&      &        &      &  101     \\
&      &        &      & 1000     \\
\hline
n &     5&       6&        7&         8 \\
\hline
&     5&       6&        7&         8  \\
&    13&      14&       15&        16  \\
&    21&      22&       23&        24  \\
&   102&      30&       31&        32  \\
&   110&     103&      104&        40  \\
&  1001&    111 &     112 &      105   \\
& 10000&    200 &     120 &      113   \\
&       &   1002 &     201 &      121  \\
&       &   1010 &    1003 &      202  \\
&       &  10001 &    1011 &      210  \\
&       &  100000&    1100 &     1004  \\                             
&       &         &  10002  &    1012  \\
&       &         &  10010  &    1020  \\
&       &         & 100001  &    1101  \\
&       &         &1000000  &    2000  \\
&       &         &         &   10003  \\
&       &         &         &   10011  \\
&       &         &         &   10100  \\
&       &         &         &  100002  \\
&       &         &         &  100010  \\
&       &         &         & 1000001  \\
&       &         &         &10000000  \\
\hline
\end{tabular}
\end{center}
\end{table}

\end{document}